\documentclass[conference]{IEEEtran}
\IEEEoverridecommandlockouts
\usepackage{cite}
\usepackage{amsmath,amssymb,amsfonts}
\usepackage{algorithmic}
\usepackage{graphicx}
\usepackage{textcomp}
\usepackage{xcolor}
\usepackage{graphicx}
\usepackage{multirow}
\usepackage{amssymb}
\usepackage{cite}
\def\BibTeX{{\rm B\kern-.05em{\sc i\kern-.025em b}\kern-.08em
    T\kern-.1667em\lower.7ex\hbox{E}\kern-.125emX}}
\begin{document}

\title{Enhancing Epileptic Seizure Detection \\with EEG Feature Embeddings}

\author{

\author{\IEEEauthorblockN{Arman Zarei\IEEEauthorrefmark{1}\IEEEauthorrefmark{2},
Bingzhao Zhu\IEEEauthorrefmark{1}\IEEEauthorrefmark{3}, and Mahsa Shoaran\IEEEauthorrefmark{1}}
\IEEEauthorblockA{azarei@umd.edu,\{bingzhao.zhu, mahsa.shoaran\}@epfl.ch}
\IEEEauthorrefmark{1}Institute of Electrical and Micro Engineering, Neuro-X Institute, EPFL, Lausanne, Switzerland\\
\IEEEauthorrefmark{2}Department of Computer Science, University of Maryland, College Park, MD, USA\\
\IEEEauthorrefmark{3}Department of Applied and Engineering Physics, Cornell University, Ithaca, NY, USA
}
}

\maketitle

\begin{abstract}
Epilepsy is one of the most prevalent brain disorders  that disrupts the lives of millions worldwide. For patients with drug-resistant seizures, there exist implantable devices capable of monitoring neural activity, promptly triggering neurostimulation to regulate seizures, or alerting patients of potential episodes. Next-generation seizure detection systems heavily rely on high-accuracy machine learning-based classifiers to detect the seizure onset. Here, we propose to enhance the seizure detection performance  by learning informative embeddings of the EEG signal. We empirically demonstrate, for the first time, that converting raw EEG signals to appropriate embeddings can significantly boost the performance of seizure detection algorithms. Importantly, we show that embedding features, which converts the raw EEG  into an alternative representation, is beneficial for various machine learning models such as Logistic Regression, Multi-Layer Perceptron,  Support Vector Machines, and  Gradient Boosted Trees. The experiments were conducted on the CHB-MIT scalp EEG dataset. With the proposed EEG feature embeddings, we achieve significant improvements in sensitivity, specificity, and AUC score across multiple models. By employing this approach alongside an SVM classifier, we were able to attain state-of-the-art classification performance  with a sensitivity of 100\% and specificity of 99\%, setting a new benchmark in the field.\newline
\end{abstract}

\section{Introduction}
A large population globally is affected by epilepsy, a neurological disorder that is recognized by the repetitive occurrence of  seizures. 
The frequency and intensity of seizures have a profound impact  on an individual's quality of life, affecting their ability to work, participate in social activities, and overall well-being. Early detection and management of seizures are crucial  to minimize their devastating effect on patients. 
Seizures are represented by episodes of high-amplitude low-frequency, or low-amplitude high-frequency neural activity in the brain \cite{schindler2007assessing, shoaran2015fully, shoaran2014compact}. 
Current seizure detection systems operate by constantly acquiring neural data from either scalp (i.e., EEG) or intracranial electrodes (i.e., iEEG), followed by signal processing to identify the onset  of seizures. This technology offers substantial benefits for individuals with epilepsy and enables them  to seek assistance promptly, mitigate the risk of injury,  or receive a closed-loop therapy (e.g.,  brain stimulation).

Today, machine learning (ML) is being used to develop predictive models for disease diagnosis and treatment in various neurological and psychiatric disorders \cite{shoeb2010application,yao2020improved,altaf201516,yao2021predicting,yao2018resting,zhu2019migraine}.
The majority of the current approaches use handcrafted EEG/iEEG biomarkers (e.g.,  band  power or time-domain features)  directly as  input to the ML models \cite{altaf201516, o2018nurip, chua2022soul,zhu2020resot}. However, the learning capabilities of such methods tend  to be limited due to spectral bias that arises during the training process, as reported in \cite{rahaman2019spectral}. Recently, \cite{tancik2020fourier}  introduced a modern solution to this problem by encoding the input space of ML models with an embedding module. The embedding module could effectively resolve these optimization complexities and enhance the learning power of  Multi-Layer Perceptrons (MLPs) as a universal approximator \cite{hornik1991approximation, cybenko1989approximation}.
Inspired by this line of work, \cite{gorishniy2022embeddings}  applied  a similar technique to general tabular datasets by mapping numerical features into feature embeddings to enhance the learning capabilities of Deep Neural Networks (DNNs). 

Building upon these influential works in computer science  \cite{gorishniy2022embeddings, tancik2020fourier}, we aimed to study   the effectiveness of  embedding approach within the realm of seizure detection for epilepsy. 
Based on our findings, we empirically demonstrate that EEG feature embeddings are highly beneficial for identifying seizure events. We tested this technique on commonly used  classifiers for seizure detection such as Logistic Regression (LR), Multi-Layer Perceptron (MLP),  Support Vector Machine (SVM), and Light Gradient Boosted Machine (LGBM).
We observed notable improvements in sensitivity, specificity, and AUC score on all the aforementioned models, using the scalp EEG dataset of the Children's Hospital Boston.
Remarkably, the utilization of feature embeddings alongside an SVM classifier yielded an unprecedented performance, achieving  an average  sensitivity of $100\%$ and specificity of $99\%$ on 24 patients. Our findings have important implications  for the development of future seizure warning systems, as well as for other studies involving neural signal classification and closed-loop stimulation. 

The remainder of this paper is organized as follows. Section~II presents the proposed feature embedding framework. Model evaluations and results are discussed in  Section \ref{experiments_section}, and Section IV concludes the paper. 

\begin{figure*}[t]
    \centering
    \includegraphics[scale=0.39]{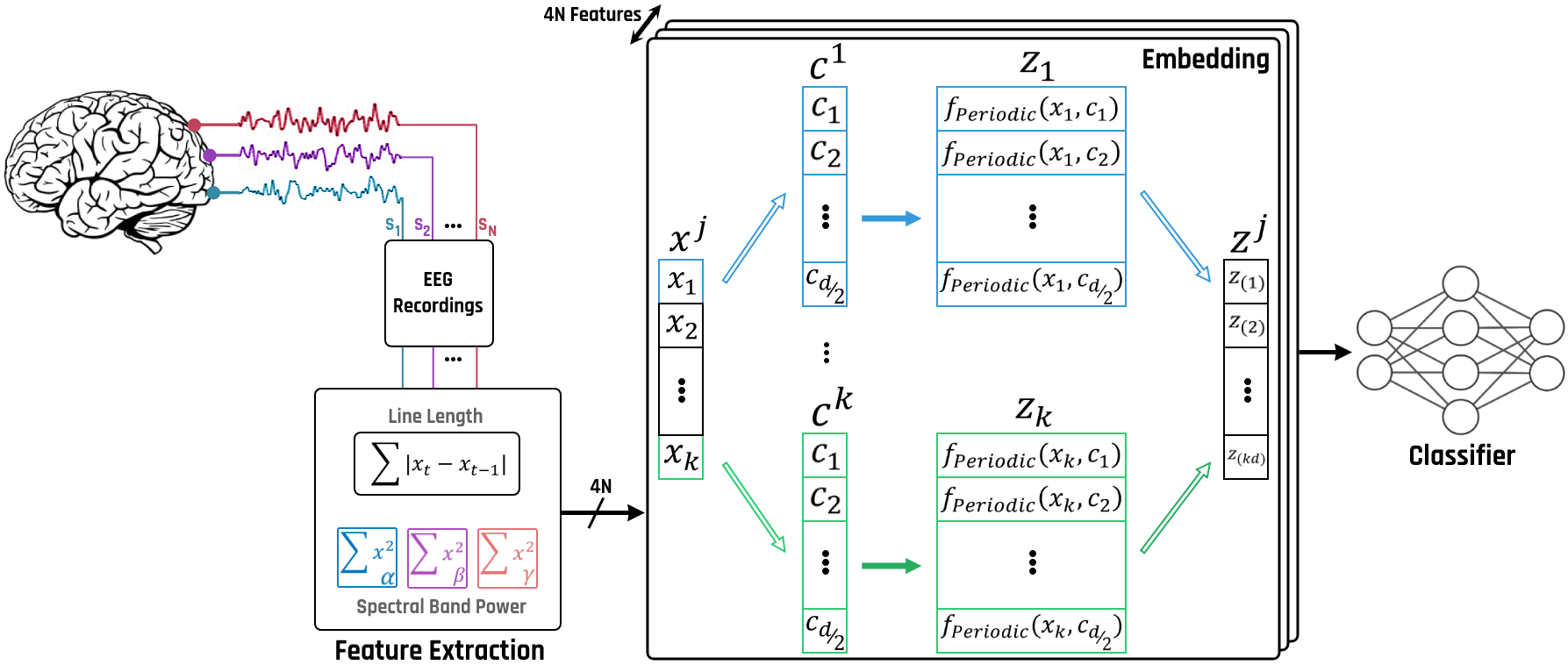}
    \caption{An overview of the proposed method that  involves collecting EEG signals $S_1, S_2, ..., S_N$ from N channels on the scalp, extracting four seizure biomarkers from each channel, and calculating their embeddings. Each feature $x_i$ is associated with an embedding function $f_{Periodic}$ with the parameter vector $c^i$ ($\{c^i_1, ... , c^i_{d/2}\}$ where the $i$ index is omitted for simplicity) to convert the scalar values into high-dimensional embeddings. The outputs are then concatenated ($z^j = concat(z_{(1)}, ..., z_{(kd)}) = concat(z_1, ...., z_k)$) and fed into the classification model to predict the probability of a seizure event.}
    \label{fig_model}
    \vspace{-3mm}
\end{figure*}

\section{Methods} \label{method}
In this Section, we present the overall framework for learning EEG feature embeddings, which can be applied to MLP and other classifiers. We extract multi-band spectral powers and line-length \cite{zhu2020resot} from 1-second EEG epochs  to construct a feature vector. Rather than using the raw features as  input to the ML models, we encode the scalar values of  EEG biomarkers and convert each numerical feature to an EEG embedding \cite{gorishniy2022embeddings}. We later show that the embedding step can lead to superior  performance in seizure detection. 

We denote the dataset as $\{(x^j, y^j)\}^n_{j=1}$ where $x^j \in \mathbb{R}^k$ represents the EEG features extracted from the $j$-th epoch 
(as discussed in detail in  Section \ref{sec_feature_extraction}) and $y^j$ denotes the binary annotation of seizure events (seizure or non-seizure). Thus, $x_i^j$ would represent the $i$-th feature of the $j$-th epoch. For simplicity, we  omit the $j$ index in the following. 
We then formalize the notion of EEG feature embeddings as:
$$z_i = f_i  (x_i) \in \mathbb{R}^{d_i}, \eqno{(1)}$$

where $z_i$ is the embedding vector, $f_i(x)$ refers to the embedding function, and $d_i$ is the dimension of the embedding space for the $i$-th feature. 
Equation (1) highlights the fact that the calculation of embeddings for each feature is performed independently. In this study, different embedding functions ($f_i$) share the same dimension in the embedding space ($d_1 = d_2 = ... = d_k$). However, the parameters of $f_i$ are not shared across different EEG biomarkers. 

After calculating $z_i$ for each $x_i$, the EEG feature embeddings are flattened and concatenated into a $1$-D vector which is passed to a classification model. More specifically, we define the EEG seizure classifier $Classifier(z)$ as:
$$
Classifier(z) = Classifier(concat(z_1, ..., z_k)), \eqno{(2)}
$$
\noindent where $k$ is the total number of features extracted from each EEG epoch, and $concat(z_1, ..., z_k) \in \mathbb{R}^{d_1 + ... + d_k}$ with $z_{1,2,...,k}$ being calculated from Equation (1).

As discussed in \cite{gorishniy2022embeddings}, there are multiple variations of $f_i$ to calculate the embedding of the input vector. Here, we employ the Periodic Activation Functions which are shown to be optimal  \cite{gorishniy2022embeddings}. Thus, we define $f_i$ as follows:
$$
f_i(x_i) = [f_{Periodic}(x_i, c^i_1), ... , f_{Periodic}(x_i, c^i_\frac{d_i}{2})], \eqno{(3)}
$$
\noindent where $d_i$ is an even number and $f_{Periodic}(x_i, c^i_l)$ is defined as: 
$$f_{Periodic}(x_i, c^i_l) = [cos(2 \pi c^i_l x_i), sin(2 \pi c^i_l x_i)], \eqno{(4)}$$

The  $c^i_l$ parameters can be either constant or variables that are trained during the model training process. In this study, we assume they are constants  initialized from a normal distribution $\mathcal{N}(0, 1)$. 
Figure \ref{fig_model} illustrates the pipeline of the proposed method. The forward pass  is initiated by the calculation of  EEG feature embeddings. These embeddings are then concatenated and fed to a machine learning classifier (e.g., LR or MLP) for predicting seizure onsets. 

\section{Results}\label{experiments_section}

In this Section, we present the experimental details of the proposed EEG feature embedding method followed by a thorough evaluation of the seizure detection performance.

\begin{table}[t]
\caption{Quantitative classification results  on 24 patients.}
\vspace{-7mm}
\label{table_results}
\begin{center}
\setlength{\tabcolsep}{2.5pt}
\renewcommand{\arraystretch}{1.4}
\begin{tabular}{c|ccc|ccc}
\multirow{2}{*}{Classifier}  & \multicolumn{3}{c|}{Conventional Approach}             & \multicolumn{3}{c}{with Embedding Module}          \\
                             & Sensitivity & Specificity & AUC    & Sensitivity  & Specificity    & AUC            \rule[-1.5ex]{0pt}{0pt} \\ \hline
\rule{0pt}{2.6ex}LR     & $94.74\%$       & $80.60\%$       & $0.8816$ & $\mathbf{100\%}$ & $\mathbf{97.29\%}$ & $\mathbf{0.9704}$ \\
MLP & $98.25\%$       & $84.95\%$       & $0.8630$ & $\mathbf{100\%}$ & $\mathbf{98.45\%}$ & $\mathbf{0.9764}$ \\
SVM & $96.49\%$       & $94.09\%$       & $0.8524$ & $\mathbf{100\%}$ & $\mathbf{99.02\%}$ & $\mathbf{0.9812}$ \\
KNN    & $96.49\%$       & $92.97\%$       & $0.8944$ & $\mathbf{100\%}$ & $\mathbf{95.33\%}$ & $\mathbf{0.9545}$ \\
GNB   & $98.25\%$       & $73.94\%$       & $0.8083$ & $\mathbf{100\%}$ & $\mathbf{89.33\%}$ & $\mathbf{0.9006}$ \\
BNB        & $50.88\%$       & $54.11\%$       & $0.4995$ & $\mathbf{100\%}$ & $\mathbf{88.88\%}$ & $\mathbf{0.9136}$ \\
LGBM        & $100\%$       & $97.29\%$       & $0.9689$ & $\mathbf{100\%}$ & $\mathbf{97.41\%}$ & $\mathbf{0.9719}$\rule[-1.5ex]{0pt}{0pt}

\end{tabular}
\end{center}
\vspace{-5mm}
\end{table}

\subsection{Dataset}
We validated our approach on the widely-used CHB-MIT \cite{goldberger2000physiobank} EEG dataset that includes 24 epileptic patients with 165 annotated seizures. We use 1-second epochs for extracting EEG biomarkers. Each epoch is annotated as either ``seizure" (i.e., $y=1$) or ``non-seizure" (i.e., $y=0$) by expert neurologists. We exclude those epochs that include both seizure and non-seizure periods. This step is crucial as it ensures that the resulting windows contain either a clear label of 1 indicating the presence of a seizure event, or a label of 0 in the absence of a seizure. Overall, we used $6805$ seizure epochs and $46917$ non-seizure epochs for training, and $3912$ seizure epochs and $890537$ non-seizure epochs for testing our models.

\subsection{Feature Extraction}\label{sec_feature_extraction}

We compute four EEG biomarkers from each channel, which are later converted to feature embeddings. Specifically, the feature extraction unit calculates the line-length (i.e., the sum of absolute  differences between consecutive samples), as well as the $\alpha$ ($8-12$ Hz), $\beta$ ($12-30$ Hz) and $\gamma$ ($30-100$ Hz)  band powers \cite{shoaran2018energy, shin2022neuraltree}. 
Multiple studies have demonstrated the effectiveness of these features in successfully discriminating between seizure and non-seizure periods \cite{altaf201516, chua2022soul, shin2022neuraltree, zhu2020resot}.

\subsection{EEG Feature Embedding}
As discussed in the Methods Section, we opted to use the Periodic Activation Function as the embedding function. We set the $d_i$ parameter (dimension of the embedding) to $20$, while  $c_i$ is sampled from a normal distribution $\mathcal{N}(0, 1)$  in all of our experiments. 
Moreover,  we use  the \textit{QuantileTransformer} (from scikit-learn) to process the EEG biomarkers \cite{gorishniy2022embeddings}, with the number of  quantiles set to $50$.
This preprocessing step  along with the Periodic module  improve the seizure detection performance. 
The hyperparameters used in the experiments were optimized for a particular patient. The same hyperparameters were then applied to the remaining patients to demonstrate the generalizability of our proposed approach. It should be noted that the performance can be enhanced further by optimizing the hyperparameters for each subject.

\begin{figure}[t]
    \centering
    \includegraphics[scale=0.18]{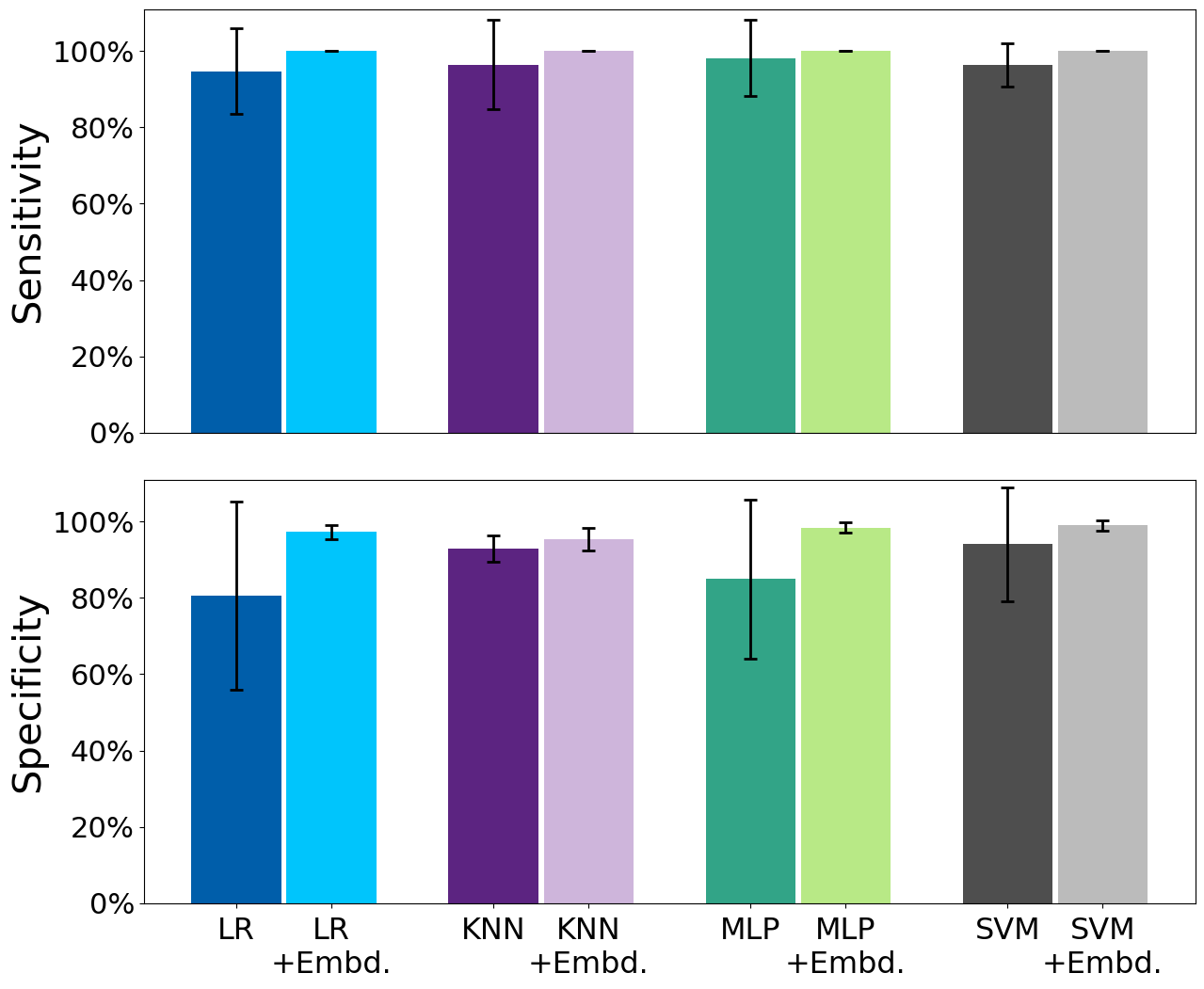}
    \caption{Comparison of average sensitivity and specificity using LR, KNN, MLP, and SVM, with or without  feature embeddings.  The use of the embedding module improves the sensitivity and specificity on all the studied models. Furthermore, the embedding module distinctly reduces the standard deviation (illustrated by  error bars) for sensitivity and specificity across patients.}
    \label{fig_avgres}
    \vspace{-3mm}
\end{figure}

\begin{table*}[t]
\caption{Comparison with the State-of-the-Art Seizure Detection Systems.}
\vspace{-4mm}
\label{table_sota}
\begin{center}
\renewcommand{\arraystretch}{1.5}
\begin{tabular}{c|c|c|c|c|c|c|c|}
\cline{2-7}
                                           & \textbf{JSSC'15 \cite{altaf201516}} & \textbf{ISSCC’20 \cite{wang2021closed}}                                             & \textbf{JSSC'22 \cite{chua2022soul}}       & \textbf{IEEE Access'20 \cite{usman2020epileptic}} & 
                                           \textbf{TNSRE'22 \cite{zhang2022epileptic}} &
                                           \textbf{This work} \\ \hline
\multicolumn{1}{|c|}{\textbf{Dataset (\# of  Patients)}}     & CHB-MIT (14)                                                     & CHB-MIT (23)     & CHB-MIT (24)  
& CHB-MIT (24)  & CHB-MIT (24)   & CHB-MIT (24)       \\ \hline
\multicolumn{1}{|c|}{\textbf{Classifier}}  & D2A-LSVM         & \begin{tabular}[c]{@{}c@{}}Coarse/Fine\\ LS-SVM\end{tabular} & LR + SGD    & CNN + SVM &   Bi-GRU    & \textbf{Embed. + SVM}    \\ \hline
\multicolumn{1}{|c|}{\textbf{Sensitivity (\%)}} & 95.7             & 97.8                                                                    & 97.5                   & 92.7    & 95.49         & \textbf{100}                \\ \hline
\multicolumn{1}{|c|}{\textbf{Specificity (\%)}} & 98.0             & 99.7                                                                     & 98.2                   & 90.8      & 98.49       & \textbf{99.0}                 \\ \hline
\end{tabular}
\end{center}
\vspace{-3mm}
\end{table*}

\subsection{ML Models} \label{models_section}
The experiments were conducted on six diverse ML models including Logistic Regression (LR), Multi-Layer Perceptron (MLP), Support Vector Machine (SVM), Gaussian Naive Bayes (GNB), Bernoulli Naive Bayes (BNB), K-Nearest Neighbors (KNN), and Light Gradient Boosting Machine (LGBM), which is an efficient implementation of the gradient boosting tree ensembles. The MLP model has two hidden layers of size $512$ and $256$, while the SVM  employs  a polynomial kernel with a degree of $6$. The LGBM was trained with $50$ estimators and a maximum depth of $5$.
The remaining model parameters were set to their default values provided in scikit-learn \cite{pedregosa2011scikit} and LightGBM library.

\begin{figure}[t]
    \centering
    \includegraphics[scale=0.45]{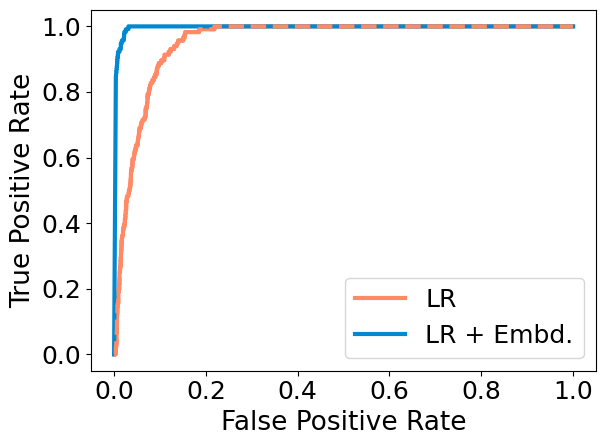} 
    \vspace{-3mm}
    \caption{Comparison of the ROC curves for logistic regression, with and without the embedding module on an arbitrary patient (\#5). LR achieves an AUC score of 0.954, whereas LR+Embd. achieves a score of 0.997, indicating the impact of having the embedding module in EEG seizure detection systems.}
    \label{fig_roc}
    \vspace{-6mm}
\end{figure}

\subsection{Seizure Detection Results}
Figure \ref{fig_avgres} shows the performance of  the widely used  ML models  for epileptic seizure detection, with and without feature embeddings. The average sensitivity and specificity of the LR are improved by $5.26\%$ and $16.69\%$, respectively, using the proposed EEG feature embeddings. This  approach is not limited to simple linear models and can be employed with more advanced algorithms such as MLP, non-linear SVM, and KNN. As demonstrated in Fig. \ref{fig_avgres}, we achieve an average improvement of $1.75\%$ in sensitivity and $13.5\%$ in  specificity, using an MLP classifier. This figure also demonstrates the improvements achieved on SVM and KNN. Furthermore, the numerical improvements in sensitivity, specificity, and AUC score for the models discussed in Section \ref{models_section}  can be seen in Table \ref{table_results}. Here, the bold numbers represent the enhanced performance achieved by the proposed feature embedding technique compared to the conventional models. We achieve an average improvement of $0.0888$, $0.1134$, and $0.1288$ in the  AUC score for the LR, MLP, and SVM, respectively. While   gradient boosted trees achieve the highest  performance prior to embedding, SVM and MLP  with embedding  outperform the LightGBM. 
Furthermore, we achieve improvements of $20.42\%$ for SVM, $3.93\%$ for MLP, $15.25\%$ for LR, $12\%$ for KNN, $19.44\%$ for GNB,  $40.28\%$ for BNB, and $0.13\%$ for LGBM in terms of epoch-based sensitivity. Since  the majority of prior work reported event-based sensitivity,
we use this measure throughout the paper. 
In addition, Fig. \ref{fig_roc} shows a comparison of the ROC curves of the LR on Patient \#5, with and without  embedding. Our analysis shows that the AUC score of Logistic Regression for this specific patient increases from 0.954 to 0.997 when used in combination with the embedding module.

\begin{figure}[t]
    \centering
    \includegraphics[scale=0.255]{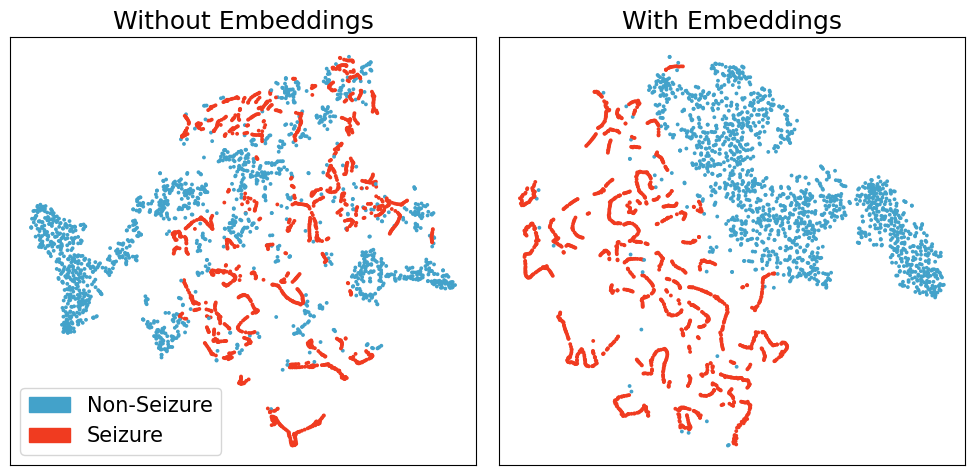} 
    \vspace{-5mm}
    \caption{Comparison of features for  patient (\#21) with and without embeddings in a 2D space, using dimensionality reduction through t-SNE. The inclusion of the embedding module  significantly improves the separability of the seizure versus non-seizure states.}
    \label{fig_tsne}
    \vspace{-5mm}
\end{figure}

Figure \ref{fig_tsne} provides a visual representation of EEG epochs for patient \#21 of the CHB-MIT dataset. Here, the t-SNE is employed to project the high-dimensional feature space onto two dimensions \cite{van2008visualizing}. Seizure epochs are shown by red dots, while blue dots represent the non-seizure epochs. It is evident that the inclusion of the proposed embedding module significantly improves the separation between seizure and non-seizure states. The improved separability  enhances the performance of ML models, thus explaining the higher performance achieved by the integration of the~embedding~module.

\begin{figure}[t]
    \centering
    \includegraphics[scale=0.2160]{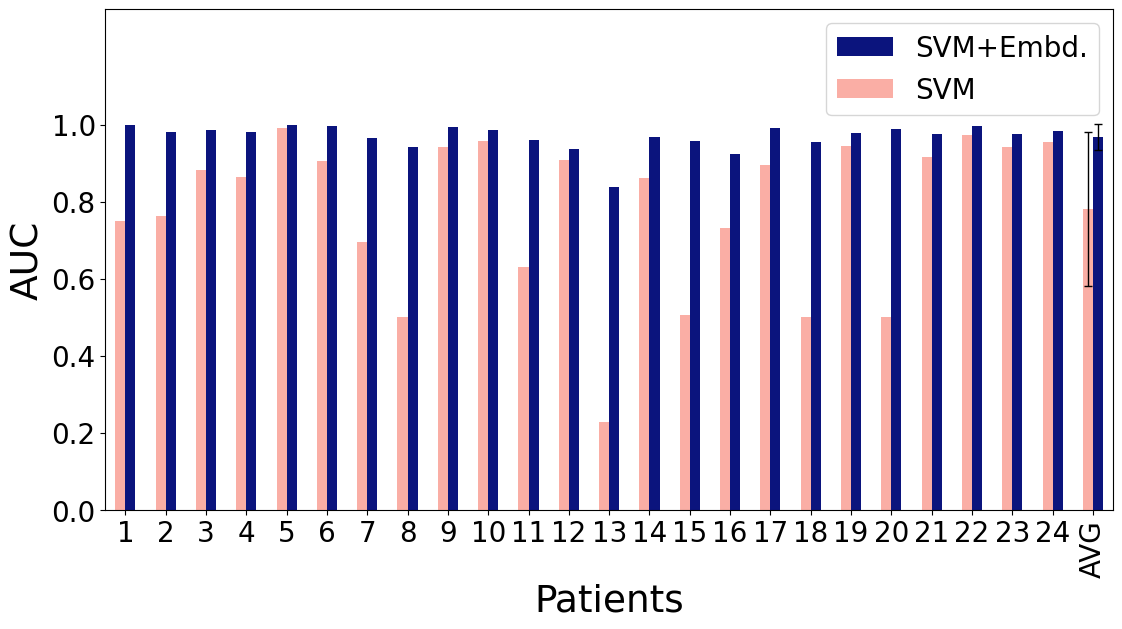} 
    \vspace{-1mm}
    \caption{Comparison of the AUC scores for SVM,  with and without the embedding module  on 24 CHB-MIT patients. 
    SVM achieves an average AUC score of $0.7879$ ($\pm0.20$), whereas SVM+Embd. achieves an enhanced score of $0.9728$ ($\pm0.03$). It can be seen that this method has improved the performance of the classifier on each individual patient in the dataset. 
    }
    \label{fig_patients_auc}
    \vspace{-3mm}
\end{figure}

Finally, we demonstrate the effectiveness of our proposed method by comparing the performance achieved by an SVM model  integrated with the embedding module, against the state-of-the-art seizure detectors. As depicted in Table \ref{table_sota}, our model achieves a superior performance, exhibiting a remarkable sensitivity of 100\% by accurately detecting all seizure events and a specificity of 99\%.  Figure \ref{fig_patients_auc} further illustrates the model's AUC score  on each patient of the CHB-MIT dataset, unequivocally highlighting the substantial performance enhancement yielded by the embedding module.

Given the universality of the proposed method (i.e., its effectiveness regardless of the ML model) and its generalizability across subjects, we plan to consider this approach for future hardware implementation. In our design, feature embeddings only require 720 parameters for a patient with 18 EEG channels (4 features per channel, embedding dimension of 20), which is significantly less compared to models such as CNN \cite{hossain2019applying}. By sharing the embedding computational module across features (e.g., utilizing a TDM approach as in \cite{shin2022neuraltree}), training the model to utilize only the top-performing features, and reducing preprocessing steps such as QuantileTransformer, 
we will improve the hardware efficiency of our approach. Nevertheless, this work serves as a compelling proof-of-concept for the potential impact of embedding techniques in enhancing the accuracy of neural signal classification.

\section{CONCLUSIONS}\vspace{-1mm}
In this work, we present, for the first time,  the concept of learning EEG feature embeddings to enhance the performance in  the epileptic seizure detection task. By evaluating the proposed model on the CHB-MIT dataset of 24  patients, we show that EEG embeddings substantially enhance the performance of seizure detection systems,  using a variety of  machine learning models. Our results demonstrate the efficacy of this method and emphasize its potential in assisting patients  with epilepsy and other neurological disorders. 
\bibliographystyle{IEEEtran}
\bibliography{ref}

\begin{thebibliography}{10}
\providecommand{\url}[1]{#1}
\csname url@samestyle\endcsname
\providecommand{\newblock}{\relax}
\providecommand{\bibinfo}[2]{#2}
\providecommand{\BIBentrySTDinterwordspacing}{\spaceskip=0pt\relax}
\providecommand{\BIBentryALTinterwordstretchfactor}{4}
\providecommand{\BIBentryALTinterwordspacing}{\spaceskip=\fontdimen2\font plus
\BIBentryALTinterwordstretchfactor\fontdimen3\font minus
  \fontdimen4\font\relax}
\providecommand{\BIBforeignlanguage}[2]{{%
\expandafter\ifx\csname l@#1\endcsname\relax
\typeout{** WARNING: IEEEtran.bst: No hyphenation pattern has been}%
\typeout{** loaded for the language `#1'. Using the pattern for}%
\typeout{** the default language instead.}%
\else
\language=\csname l@#1\endcsname
\fi
#2}}
\providecommand{\BIBdecl}{\relax}
\BIBdecl

\bibitem{schindler2007assessing}
K.~Schindler, H.~Leung, C.~E. Elger, and K.~Lehnertz, ``Assessing seizure
  dynamics by analysing the correlation structure of multichannel intracranial
  eeg,'' \emph{Brain}, vol. 130, no.~1, pp. 65--77, 2007.

\bibitem{shoaran2015fully}
M.~Shoaran, C.~Pollo, K.~Schindler, and A.~Schmid, ``A fully integrated ic with
  0.85-$\mu$w/channel consumption for epileptic ieeg detection,'' \emph{IEEE
  Transactions on Circuits and Systems II: Express Briefs}, vol.~62, no.~2, pp.
  114--118, 2015.

\bibitem{shoaran2014compact}
M.~Shoaran, M.~H. Kamal, C.~Pollo, P.~Vandergheynst, and A.~Schmid, ``Compact
  low-power cortical recording architecture for compressive multichannel data
  acquisition,'' \emph{IEEE transactions on biomedical circuits and systems},
  vol.~8, no.~6, pp. 857--870, 2014.

\bibitem{shoeb2010application}
A.~H. Shoeb and J.~V. Guttag, ``Application of machine learning to epileptic
  seizure detection,'' in \emph{Proceedings of the 27th international
  conference on machine learning (ICML-10)}, 2010, pp. 975--982.

\bibitem{yao2020improved}
L.~Yao, P.~Brown, and M.~Shoaran, ``Improved detection of parkinsonian resting
  tremor with feature engineering and kalman filtering,'' \emph{Clinical
  Neurophysiology}, vol. 131, no.~1, pp. 274--284, 2020.

\bibitem{altaf201516}
M.~A. Bin~Altaf, C.~Zhang, and J.~Yoo, ``A 16-channel patient-specific seizure
  onset and termination detection soc with impedance-adaptive transcranial
  electrical stimulator,'' \emph{IEEE Journal of Solid-State Circuits},
  vol.~50, no.~11, pp. 2728--2740, 2015.

\bibitem{yao2021predicting}
L.~Yao, J.~L. Baker, N.~D. Schiff, K.~P. Purpura, and M.~Shoaran, ``Predicting
  task performance from biomarkers of mental fatigue in global brain
  activity,'' \emph{Journal of neural engineering}, vol.~18, no.~3, p. 036001,
  2021.

\bibitem{yao2018resting}
L.~Yao, P.~Brown, and M.~Shoaran, ``Resting tremor detection in parkinson's
  disease with machine learning and kalman filtering,'' in \emph{2018 IEEE
  Biomedical Circuits and Systems Conference (BioCAS)}.\hskip 1em plus 0.5em
  minus 0.4em\relax IEEE, 2018, pp. 1--4.

\bibitem{zhu2019migraine}
B.~Zhu, G.~Coppola, and M.~Shoaran, ``Migraine classification using
  somatosensory evoked potentials,'' \emph{Cephalalgia}, vol.~39, no.~9, pp.
  1143--1155, 2019.

\bibitem{o2018nurip}
G.~O’Leary, D.~M. Groppe, T.~A. Valiante, N.~Verma, and R.~Genov, ``Nurip:
  Neural interface processor for brain-state classification and
  programmable-waveform neurostimulation,'' \emph{IEEE Journal of Solid-State
  Circuits}, vol.~53, no.~11, pp. 3150--3162, 2018.

\bibitem{chua2022soul}
A.~Chua, M.~I. Jordan, and R.~Muller, ``Soul: An energy-efficient unsupervised
  online learning seizure detection classifier,'' \emph{IEEE Journal of
  Solid-State Circuits}, vol.~57, no.~8, pp. 2532--2544, 2022.

\bibitem{zhu2020resot}
B.~Zhu, M.~Farivar, and M.~Shoaran, ``Resot: Resource-efficient oblique trees
  for neural signal classification,'' \emph{IEEE Transactions on Biomedical
  Circuits and Systems}, vol.~14, no.~4, pp. 692--704, 2020.

\bibitem{rahaman2019spectral}
N.~Rahaman, A.~Baratin, D.~Arpit, F.~Draxler, M.~Lin, F.~Hamprecht, Y.~Bengio,
  and A.~Courville, ``On the spectral bias of neural networks,'' in
  \emph{International Conference on Machine Learning}.\hskip 1em plus 0.5em
  minus 0.4em\relax PMLR, 2019, pp. 5301--5310.

\bibitem{tancik2020fourier}
M.~Tancik, P.~Srinivasan, B.~Mildenhall, S.~Fridovich-Keil, N.~Raghavan,
  U.~Singhal, R.~Ramamoorthi, J.~Barron, and R.~Ng, ``Fourier features let
  networks learn high frequency functions in low dimensional domains,''
  \emph{Advances in Neural Information Processing Systems}, vol.~33, pp.
  7537--7547, 2020.

\bibitem{hornik1991approximation}
K.~Hornik, ``Approximation capabilities of multilayer feedforward networks,''
  \emph{Neural networks}, vol.~4, no.~2, pp. 251--257, 1991.

\bibitem{cybenko1989approximation}
G.~Cybenko, ``Approximation by superpositions of a sigmoidal function,''
  \emph{Mathematics of control, signals and systems}, vol.~2, no.~4, pp.
  303--314, 1989.

\bibitem{gorishniy2022embeddings}
Y.~Gorishniy, I.~Rubachev, and A.~Babenko, ``On embeddings for numerical
  features in tabular deep learning,'' \emph{Advances in Neural Information
  Processing Systems}, vol.~35, pp. 24\,991--25\,004, 2022.

\bibitem{goldberger2000physiobank}
A.~L. Goldberger, L.~A. Amaral, L.~Glass, J.~M. Hausdorff, P.~C. Ivanov, R.~G.
  Mark, J.~E. Mietus, G.~B. Moody, C.-K. Peng, and H.~E. Stanley, ``Physiobank,
  physiotoolkit, and physionet: components of a new research resource for
  complex physiologic signals,'' \emph{circulation}, vol. 101, no.~23, pp.
  e215--e220, 2000.

\bibitem{shoaran2018energy}
M.~Shoaran, B.~A. Haghi, M.~Taghavi, M.~Farivar, and A.~Emami-Neyestanak,
  ``Energy-efficient classification for resource-constrained biomedical
  applications,'' \emph{IEEE Journal on Emerging and Selected Topics in
  Circuits and Systems}, vol.~8, no.~4, pp. 693--707, 2018.

\bibitem{shin2022neuraltree}
U.~Shin, C.~Ding, B.~Zhu, Y.~Vyza, A.~Trouillet, E.~C.~M. Revol, S.~P. Lacour,
  and M.~Shoaran, ``Neuraltree: A 256-channel 0.227-$\mu$j/class versatile
  neural activity classification and closed-loop neuromodulation soc,''
  \emph{IEEE Journal of Solid-State Circuits}, vol.~57, no.~11, pp. 3243--3257,
  2022.

\bibitem{wang2021closed}
Y.~Wang, Q.~Sun, H.~Luo, X.~Chen, X.~Wang, and H.~Zhang, ``A closed-loop
  neuromodulation chipset with 2-level classification achieving 1.5vpp cm
  interference tolerance, 35db stimulation artifact rejection in 0.5ms and 97.8
  sensitivity seizure detection,'' in \emph{2020 IEEE International Solid-
  State Circuits Conference - (ISSCC)}, 2020, pp. 406--408.

\bibitem{usman2020epileptic}
S.~M. Usman, S.~Khalid, and M.~H. Aslam, ``Epileptic seizures prediction using
  deep learning techniques,'' \emph{Ieee Access}, vol.~8, pp. 39\,998--40\,007,
  2020.

\bibitem{zhang2022epileptic}
Y.~Zhang, S.~Yao, R.~Yang, X.~Liu, W.~Qiu, L.~Han, W.~Zhou, and W.~Shang,
  ``Epileptic seizure detection based on bidirectional gated recurrent unit
  network,'' \emph{IEEE Transactions on Neural Systems and Rehabilitation
  Engineering}, vol.~30, pp. 135--145, 2022.

\bibitem{pedregosa2011scikit}
F.~Pedregosa, G.~Varoquaux, A.~Gramfort, V.~Michel, B.~Thirion, O.~Grisel,
  M.~Blondel, P.~Prettenhofer, R.~Weiss, V.~Dubourg \emph{et~al.},
  ``Scikit-learn: Machine learning in python,'' \emph{the Journal of machine
  Learning research}, vol.~12, pp. 2825--2830, 2011.

\bibitem{van2008visualizing}
L.~Van~der Maaten and G.~Hinton, ``Visualizing data using t-sne.''
  \emph{Journal of machine learning research}, vol.~9, no.~11, 2008.

\bibitem{hossain2019applying}
M.~S. Hossain, S.~U. Amin, M.~Alsulaiman, and G.~Muhammad, ``Applying deep
  learning for epilepsy seizure detection and brain mapping visualization,''
  \emph{ACM Transactions on Multimedia Computing, Communications, and
  Applications (TOMM)}, vol.~15, no.~1s, pp. 1--17, 2019.

\end{thebibliography}

\end{document}